# Density Evolution, Thresholds and the Stability Condition for Non-binary LDPC Codes

Vishwambhar Rathi and Ruediger Urbanke

EPFL, CH-1015 Lausanne

**Abstract**

We derive the density evolution equations for non-binary low-density parity-check (LDPC) ensembles when transmission takes place over the binary erasure channel. We introduce ensembles defined with respect to the general linear group over the binary field. For these ensembles the density evolution equations can be written compactly. The density evolution for the general linear group helps us in understanding the density evolution for codes defined with respect to finite fields. We compute thresholds for different alphabet sizes for various LDPC ensembles. Surprisingly, the threshold is not a monotonic function of the alphabet size. We state the stability condition for non-binary LDPC ensembles over any binary memoryless symmetric channel. We also give upper bounds on the MAP thresholds for various non-binary ensembles based on EXIT curves and the area theorem.

# 1 Introduction

It is well known that using binary LDPC ensembles for transmission over *binary memoryless symmetric* (BMS) channels, one can construct codes which achieve rates seemingly arbitrarily close to capacity. In particular, for the *binary erasure channel* (BEC) there are provable capacity-achieving degree distributions obtained in [7, 8, 21] based on the method of *density evolution*. The method of density evolution was generalized to any BMS channel in [18]. This generalization made it possible to construct codes for a given BMS channel which can achieve rates very close to the capacity [2, 16]. It should be noted however that good LDPC codes for a BMS channel other than the BEC are obtained by numerical optimization. The problem of finding explicitly capacity-achieving degree distribution for a general BMS channel is still open. In either case, the main property of these capacity achieving/approaching degree distributions is that the underlying parity-check matrix gets denser and denser as the gap to capacity is reduced [20].

Although *binary* ensembles are conjectured to constitute a powerful enough class to achieve capacity, it is nevertheless worth exploring the potential of non-binary ensembles. Clearly, by considering non-binary alphabets we add one more degree of freedom in our code design. Whereas the standard approach is to fix the alphabet size (to binary) and to increase the density of the underlying graph, let us take an alternative route here. Suppose we fix the degree distribution and let the alphabet size increase. As we will see shortly, if we consider the corresponding underlying binary graph, this graph becomes denser and denser as well (but of course we consider BP decoding on the non-binary graph). One might therefore hope to find better and better performance, and an increase in the alphabet size might yet yield another way of achieving capacity. The relationship between alphabet size and performance is not a simple one and an increase in the underlying alphabet does not necessarily lead to increased performance. Nevertheless, there are many unexplored degrees of freedom in the system design and this paper is only the very first step in a systematic study of these relationships.

The possibility of using non-binary alphabets for LDPC codes was already proposed by Gallager in his landmark PhD thesis [5]. The fact that by using non-binary alphabets, the performance of LDPC codes over BMS channels can be improved was first reported by Davey and MacKay [3]. They showed by specific examples that non-binary LDPC codes can perform significantly better than their binary counterparts for the BMS channels. Hu showed that even the performance of cycle codes can be improved considerably with non binary alphabets [6]. In [22], Sridhara and Fuja have designed codes over certain rings and groups for coded modulation based on the principle of non-binary LDPC codes. Also, the design of LDPC code construction using "liftings" of multi-edge type designs [15, 19] and the related framework of protographs [23] can be seen as non-binary LDPC ensembles.

Despite of these results there has been no systematic study of non-binary LDPC ensembles. In particular, no efficient method of evaluating their asymptotic performance is known. The difficulty is that the messages of the belief propagation decoder "live" in a high-dimensional space, so that it is in general difficult to keep track of their densities.

The paper is organized in the following way: in Section 2, we define various quantities of interest. Section 3 describes the message-passing decoder for non-binary alphabets. In Section 4, we derive the density evolution equations for ensembles over the general linear group and finite fields and state the stability condition. We calculate an upper bound on the MAP threshold for non-binary alphabets in Section 5. We conclude in Section 6. Because of space limitations we skip all proofs.

# 2 Preliminaries

We consider transmission over a BMS channel using non-binary LDPC ensembles. We denote the set of symbols of the codeword by $\mathcal{S}$ and its cardinality by $|\mathcal{S}| = q = 2^m$. For convenience we assume hereby that the alphabet size is a power of 2. Thus we can think of each symbol as a binary $m$-tuple. In order to transmit a symbol over a BMS channel we transmit the bits representing this symbol. We define the non-binary LDPC ensemble in an analogous way as in the binary case [16]. We define an ensemble of bipartite graphs $\mathbb{G}(n,\lambda,\rho)$ with degree distribution $(\lambda,\rho)$ and blocklength $n$. Next we assign a bijective linear mapping $f : \mathcal{S} \mapsto \mathcal{S}$ to each edge of every bipartite graph in the ensemble $\mathbb{G}(n,\lambda,\rho)$. The mappings are chosen uniformly at random from a set of mappings $\mathcal{F}$. A bipartite graph from the ensemble represents parity-check equations of the form

$$\sum_i f_i(x_i) = 0, \tag{1}$$

where the $\{x_i\}, x_i \in \mathcal{S}$, are the variables which participate in the parity check equation and $f_i \in \mathcal{F}$. Note that a code defined by the parity-check equations of the form in (1) is linear as the mappings $f$ are linear. The *design rate* of an ensemble with d.d. $(\lambda,\rho)$ is the same as in the binary case:

$$r = 1 - \frac{\int_0^1 \rho(x)dx}{\int_0^1 \lambda(x)dx}.$$

Let $x = (x_1, \cdots x_n)$ denote a codeword over $\mathcal{S}$. We can think of each symbol as a binary $m$-tuple. Hence, we can equivalently think of the codeword as a binary codeword of length $nm$, $x = (x_{11}, \cdots, x_{1m}, \cdots, x_{n1}, \cdots, x_{nm})$. To transmit the codeword $x$ over a BMS channel, we transmit its binary components and let the corresponding received word be $y = (y_{11}, \cdots, y_{1m}, \cdots, y_{n1}, \cdots, y_{nm})$.

In this paper we will consider two variants of non-binary LDPC ensembles: Ensembles over finite fields and ensembles over the general linear group. For ensembles over finite fields, $\mathcal{S} = \mathrm{GF}(2^m)$, and the mappings $f$ are of the form $f(x) = \omega x$, where $\omega \in \mathrm{GF}^*(2^m)$, the multiplicative group of $\mathrm{GF}(2^m)$. Hence, by some abuse of notation $\mathcal{F} = \mathrm{GF}^*(2^m)$. This implies that $|\mathcal{F}| = 2^m - 1$. We will denote an LDPC ensemble over $\mathrm{GF}(2^m)$ with d.d. $(\lambda, \rho)$ by $\mathrm{EGF}(\lambda, \rho, m)$ (we do not show the dependence on the block length as we will be only interested in the asymptotic limit). Note that we can find an equivalent binary code corresponding to every code in the ensemble $\mathrm{EGF}(\lambda, \rho, m)$. For example, Fig. 1 shows the Forney Style Factor Graph (FSFG [4]) of a simple code over GF(4) together with its corresponding parity-check matrix. This is equivalent to the binary code shown in Fig. 2. E.g., the constraint over GF(4), $x_1 + (1+z)x_2 = 0$ is equivalent to the two binary constraints $x_{11} + x_{21} + x_{22} = 0$ and $x_{12} + x_{22} = 0$ and so on. The corresponding binary FSFG and the binary parity-check matrix are shown in Fig. 2.

The ensemble over the general linear group is denoted by $\mathrm{EGL}(\lambda, \rho, m)$. It has as symbol set $\mathcal{S}$ the vector space $\mathrm{GF}_2^m$ of dimension $m$ over the binary field. The mappings are given by $f(b) = Wb$, where $b \in \mathrm{GF}_2^m$ and $W \in \mathrm{GL}_2^m$. $\mathrm{GL}_2^m$ is the set of all $m \times m$ invertible matrices over the binary field. Note that the number of distinct invertible matrices over the binary field is $\prod_{l=0}^{m-1} (2^m - 2^l)$, [10]. Again, we can find an equivalent binary code corresponding to a code defined with respect to the general linear group. For example, if the parity-check matrix over $\mathrm{GL}_2^2$ is given by:

$$H = \begin{pmatrix} H_{1,1} & H_{1,2} & 0 & 0 \\ 0 & 0 & H_{2,3} & H_{2,4} \\ H_{3,1} & 0 & 0 & H_{3,4} \end{pmatrix}$$

where,

$$H_{1,1} = \begin{pmatrix} 1 & 0 \\ 0 & 1 \end{pmatrix}, \quad H_{1,2} = \begin{pmatrix} 1 & 1 \\ 1 & 0 \end{pmatrix}, \quad H_{2,3} = \begin{pmatrix} 1 & 1 \\ 0 & 1 \end{pmatrix},$$

$$H_{2,4} = \begin{pmatrix} 1 & 0 \\ 1 & 1 \end{pmatrix}, \quad H_{3,1} = \begin{pmatrix} 0 & 1 \\ 1 & 1 \end{pmatrix}, \quad H_{3,4} = \begin{pmatrix} 0 & 1 \\ 1 & 0 \end{pmatrix}.$$

Then the equivalent binary matrix is given by:

$$H_b = \begin{pmatrix} 1 & 0 & 1 & 1 & 0 & 0 & 0 & 0 \\ 0 & 1 & 1 & 0 & 0 & 0 & 0 & 0 \\ 0 & 0 & 0 & 0 & 1 & 1 & 1 & 0 \\ 0 & 0 & 0 & 0 & 0 & 1 & 1 & 1 \\ 0 & 1 & 0 & 0 & 0 & 0 & 0 & 1 \\ 1 & 1 & 0 & 0 & 0 & 0 & 1 & 0 \end{pmatrix}.$$

Each constraint in the matrix $H$ is equivalent to two constraints in the equivalent binary matrix $H_b$. E.g., the constraint $H_{1,1}x_1 + H_{1,2}x_2 = 0$ is equivalent to $x_{11} + x_{21} + x_{22} = 0$ and $x_{12} + x_{21} = 0$.

We will be interested in the number of different subspaces of dimension $k$ of the vector space $\mathrm{GF}_2^m$. This number is known as the Gaussian binomial coefficient. We denote it by $\begin{bmatrix} m \\ k \end{bmatrix}$, and it is given by (see [10], pp. 443):

$$\begin{bmatrix} m \\ k \end{bmatrix} = \begin{cases} 1, & \text{if} \quad k = 0 \;\text{ or }\; k = m, \\ \prod_{l=0}^{k-1} \frac{2^m - 2^l}{2^k - 2^l}, & \text{otherwise}. \end{cases}$$

The messages in the belief propagation decoder are vectors of length $|\mathcal{S}|$. The $\alpha^{\text{th}}$ component of a message $\Psi$, denoted by $\psi(\alpha)$ where $\alpha \in \mathcal{S}$, is equal to the posteriori probability that the symbol is $\alpha$. As we are assuming transmission over the BEC, the messages have a very specific form. We will interpret the messages arising in the belief propagation decoder when transmission is over BEC as subspaces of the vector space $\mathrm{GF}_2^m$ (Lemma 4.2). We say that the dimension of a message $\Psi$, call it $\dim(\Psi)$, is $k$ if the number of non-zero entries of $\Psi$ is $2^k$. This means that $2^k$ of the $|\mathcal{S}|$ symbols have non-zero probability and in fact because of the special nature of the BEC they all have the same probability. We associate to a message $\Psi$ of dimension $k$, an $(m-k) \times m$ matrix $M$. Again because of the special nature of the message space for BEC and the message processing rules, the symbols with non-zero entries in the message $\Psi$ belongs to the null space of the matrix $M$. We denote the subspace of the non-zero entries of the message $\Psi$ by V. We say that V is the subspace of $\Psi$ and denote the dimension of V by $\dim(\mathrm{V}) = \dim(\Psi)$. The orthogonal complement of V is denoted by $\mathrm{V}^{\perp}$. We have the following relations,

$$\mathrm{V} = \mathrm{nullspace}\,(M)\,, \qquad \mathrm{V}^{\perp} = \mathrm{rowspace}\,(M)\,.$$

To illustrate this, we consider an example for $m = 2$. Lets assume that we have a variable node $x_i = \{x_{i1}, x_{i2}\}$. For the sake of simplicity assume that $x_{i1} = 0, x_{i2} = 0$ and after transmission over the BEC $x_{i1}$ is not erased and $x_{i2}$ is erased. Then the initial message from the variable node $x_i$ is $\Psi^{(1)} = \left\{\frac{1}{2}, 0, \frac{1}{2}, 0\right\}$ and $\dim(\Psi) = 1$. As the non-zero entries of $\Psi^{(1)}$ satisfies $x_{i1=0}$, so the associated matrix $M = [1 \quad 0]$. The subspace V is $\{(0,0),(0,1)\}$. Clearly $\mathrm{V} = \mathrm{nullspace}\,(M)$.

If a subspace V has the basis vectors $B = \{b_1, \ldots, b_k\}$, then we denote the set of basis vectors for $\mathrm{V}^{\perp}$ by $B^{\perp} = \{b_1^{\perp}, \ldots, b_k^{\perp}\}$. We denote the linear span of a set of vectors $B$ by $\mathrm{Span}\,(B)$.

# 3 Belief Propagation Algorithm

The messages in the belief propagation algorithm for non-binary LDPC codes from both the ensembles $\mathrm{EGF}\,(\lambda, \rho, m)$ and $\mathrm{EGL}\,(\lambda, \rho, m)$ are vectors of length $q = 2^m$. An iteration of the message passing algorithm consists of the following steps:

1. **Initial Message:** The initial message $\Psi_{x,\mathtt{c}}^{(1)}$ from a variable node $x$ to a connected check node $\mathtt{c}$ is the posteriori probability distribution of symbols $(\mathrm{P}(x = \alpha_0|y), \cdots, \mathrm{P}(x = \alpha_{q-1}|y))$, where $\alpha_i \in \mathcal{S}$, $i = 0, \ldots, q-1$.

2. **Edge Action:** Before the messages reach the check nodes, we need to consider the permutations induces by the edge labels and their associated mapping $f(x)$. Note that labels induce permutations since the mappings $f$ are invertible. More precisely, if the edge label is $f$ then the message vector $\Psi_{x,\mathtt{c}}^{(1)} = \left(\psi_{x,\mathtt{c}}^{(1)}(\alpha_0), \cdots, \psi_{x,\mathtt{c}}^{(1)}(\alpha_{q-1})\right)$ gets permuted to the message $\Psi_{x,\mathtt{c}}^{(2)} = \left(\psi_{x,\mathtt{c}}^{(1)}\left(f^{-1}(\alpha_0)\right), \cdots, \psi_{x,\mathtt{c}}^{(1)}\left(f^{-1}(\alpha_{q-1})\right)\right)$.

3. **Check Node Action:** The operation on the check node side is the convolution of the incoming messages. Lets consider a check node $\mathtt{c}$. Let its degree be 3 for the sake of simplicity. Let $x$, $y$, and $z$ be the connected variables and lets consider the outgoing message along the edge to $z$ as a function of the incoming messages along the edges connected to $x$ and $y$. The outgoing message towards variable $z$ is then

$$\begin{aligned}\psi^{(3)}_{\mathtt{c},z}(\alpha) &= \sum_{\beta,\gamma\in\mathcal{S}} \mathbb{1}_{\{\alpha+\beta+\gamma=0\}}\psi^{(2)}_{x,\mathtt{c}}(\beta)\psi^{(2)}_{y,\mathtt{c}}(\gamma) \\ &= \sum_{\beta\in\mathcal{S}} \psi^{(2)}_{x,\mathtt{c}}(\beta)\psi^{(2)}_{y,\mathtt{c}}(-\alpha-\beta), \qquad (2)\end{aligned}$$

where,

$$\mathbb{1}_{\{\delta=0\}} = \begin{cases} 1, & \text{if} \quad \delta = 0, \\ 0, & \text{otherwise.} \end{cases}$$

In a brute force manner, the above summation can be accomplished with complexity $O(q^2)$. However, note that $\psi^{(3)}_{\mathtt{c},z}(-\alpha)$ is given in terms of a *convolution* of two message vectors, where the index calculations are done with respect to the additive group of $\mathrm{GF}(2^m)$. Note that the vector space $\mathrm{GF}_2^m$ and finite field $\mathrm{GF}(2^m)$ are isomorphic groups with respect to addition. Hence, as suggested in [9, 18], we can use Fourier transforms to accomplish this convolution in an efficient manner. Since the message size is $2^m$, the Fourier transform is particularly simple. Write an element of $\mathcal{S}$ as an $m$-tuple with components in GF(2), $\alpha = (\alpha_1, \cdots, \alpha_m)$. Let $\Psi = \left(\psi(\alpha)_{\alpha\in\mathcal{S}}\right)$ denote a vector whose components are taking values in $\mathbb{C}$. Let $\Phi = (\phi(\alpha))_{\alpha\in\mathcal{S}}$ denote its Fourier transform. The corresponding Fourier transform pair is

$$\phi(\alpha) = \sum_{\beta}\psi(\beta)(-1)^{-\alpha.\beta^T}, \qquad \psi(\alpha) = \frac{1}{2^m}\sum_{\beta}\phi(\beta)(-1)^{\alpha.\beta^T}, \qquad (3)$$

where $\alpha.\beta^T$ is the dot product of the binary representation of $\alpha$ and $\beta$. Thus the check node operation is given by the component wise multiplication of the Fourier transform of incoming messages and by taking the inverse Fourier transform of the result.

4. **Inverse Edge Action:** Messages are again permuted to take care of the mappings. But in this case the permutation is the inverse of the permutation in step 2. More precisely, if the edge label is $f$ then the message vector $\Psi^{(3)}_{\mathtt{c},x}$ gets permuted to the message $\Psi^{(4)}_{\mathtt{c},x} = \left(\psi^{(3)}_{\mathtt{c},x}\left(f(\alpha_0)\right), \cdots, \psi^{(3)}_{\mathtt{c},x}\left(f(\alpha_{q-1})\right)\right)$.

5. **Variable Node Action:** The operation on the variable node side is the component-wise multiplication of the incoming messages and the initial message. We normalize the result of the multiplication.

# 4 Density Evolution For BEC

It can be shown that the concentration of the error probability holds also for the non-binary LDPC ensembles. The proof is essentially the same as in [18]. Hence, in the asymptotic limit the average behavior of the iterative decoder determines the performance of a randomly chosen code with probability one. As in [18] we can again show that the all-zero codeword assumption holds, i.e., the error probability is the same for the belief propagation decoder in the cases when the all-zero codeword is transmitted and when any other codeword is transmitted.

**Lemma 4.1 (All-Zero Codeword Assumption)** *Consider transmission over a BMS channel using an element of EGL $(\lambda, \rho, m)$ or EGF $(\lambda, \rho, m)$. Then the conditional error probability of the message passing decoder is independent of the transmitted codeword.*

## 4.1 Density Evolution for EGL $(\lambda, \rho, m)$

In order to derive the density evolution equations for the ensemble EGL $(\lambda, \rho, m)$ we need to find the set of messages which arise in the belief propagation decoder. In the following lemma we characterize all the messages which appear in the belief propagation decoder.

**Lemma 4.2 (Message Space Characterization)** *Consider the ensemble EGL $(\lambda, \rho, m)$ and transmission over the BEC. The messages arising in the belief propagation decoder satisfy the following properties:*

1. *All the non-zero entries in a message $\Psi$ are equal.*

2. *Let $V = \{\alpha \in \mathrm{GF}_2^m : \psi(\alpha) \neq 0\}$. Then $V$ is subspace of* $\mathrm{GF}_2^m$.

3. *The Fourier transform $\Phi$ of a message $\Psi$ has the property:*

$$\phi(\alpha) = \begin{cases} 1, & \text{if } \alpha \in V^{\perp}, \\ 0, & \text{otherwise}, \end{cases}$$

*where $V^{\perp}$ is the orthogonal complement of V. Thus the total number of messages is equal to $\sum_{i=0}^{m} \begin{bmatrix} m \\ i \end{bmatrix}$.*

We observe that after the edge action and the inverse edge action, all the messages of the same dimension have equal probability. Hence we only need to keep track of the probability of the dimension of a message. Thus by this observation and Lemma 4.2, we can write the density evolution as an $m+1$ dimensional recursion.

**Lemma 4.3 (Density Evolution for EGL$(\lambda, \rho, m)$)** *Consider the non-binary LDPC ensemble EGL$(\lambda, \rho, m)$. Let $\mathrm{P}_v^{(l)}(k, \mathtt{l})$ be the probability that a randomly chosen message is of dimension k after the edge action connected to a variable node of degree $\mathtt{l}$ (i.e., a message just before the check node processing). Similarly, $\mathrm{P}_c^{(l)}(k, \mathtt{r})$ denotes the probability that a randomly chosen message is of dimension k after the inverse edge action connected to a check node of degree $\mathtt{r}$ (i.e., a message just before the variable node processing). Then we have the following recursive relationships between different probabilities on the check node side:*

$$\mathrm{P}_c^{(l)}(k,3) = \sum_{i=0}^{k} \mathrm{P}_v^{(l)}(i) \sum_{j=k-i}^{k} \frac{\begin{bmatrix} m-i \\ m-k \end{bmatrix} \begin{bmatrix} i \\ k-j \end{bmatrix} 2^{(k-i)(k-j)}}{\begin{bmatrix} m \\ m-j \end{bmatrix}} \mathrm{P}_v^{(l)}(j) \tag{4}$$

$$\mathrm{P}_c^{(l)}(k,\mathtt{r}) = \sum_{i=0}^{k} \mathrm{P}_c^{(l)}(i,\mathtt{r}-1) \sum_{j=k-i}^{k} \frac{\begin{bmatrix} m-i \\ m-k \end{bmatrix} \begin{bmatrix} i \\ k-j \end{bmatrix} 2^{(k-i)(k-j)}}{\begin{bmatrix} m \\ m-j \end{bmatrix}} \mathrm{P}_v^{(l)}(j) \tag{5}$$

*where* $\mathrm{P}_v^{(l)}(i)$ *is the average over the variable node degree distribution,*

$$\mathrm{P}_v^{(l)}(i) = \sum_{\mathtt{l}} \lambda_{\mathtt{l}} \mathrm{P}_v^{(l)}(i,\mathtt{l}).$$

*The equations on the variable node side for the probabilities in* $(l+1)^{th}$ *iteration are:*

$$\mathrm{P}_v^{(l+1)}(k,2) = \sum_{i=k}^{m} \binom{m}{i} \epsilon^i (1-\epsilon)^{m-i} \sum_{j=k}^{m-i+k} \frac{\begin{bmatrix} i \\ k \end{bmatrix} \begin{bmatrix} m-i \\ j-k \end{bmatrix} 2^{(i-k)(j-k)}}{\begin{bmatrix} m \\ j \end{bmatrix}} \mathrm{P}_c^{(l)}(j) \tag{6}$$

$$\mathrm{P}_v^{(l+1)}(k,\mathtt{l}) = \sum_{i=k}^{m} \mathrm{P}_v^{(l+1)}(i,\mathtt{l}-1) \sum_{j=k}^{(m-i+k)} \frac{\begin{bmatrix} i \\ k \end{bmatrix} \begin{bmatrix} m-i \\ j-k \end{bmatrix} 2^{(i-k)(j-k)}}{\begin{bmatrix} m \\ j \end{bmatrix}} \mathrm{P}_c^{(l)}(j) \tag{7}$$

*where* $\mathrm{P}_c^{(l)}(j)$ *is the average over the variable node degree distribution,*

$$\mathrm{P}_c^{(l)}(j) = \sum_{\mathtt{r}} \rho_{\mathtt{r}} \mathrm{P}_c^{(l)}(j,\mathtt{r}).$$

In Table 4.1, we list the thresholds for various ensembles. Note that for the ensemble with d.d. pair $\lambda(y) = y, \rho(y) = y^2$, initially the threshold increases rapidly (as $m$ is increased). Unfortunately it reaches a peak at $m = 6$ and then starts decreasing. For the ensemble with d.d. pair $\lambda(y) = 0.5y + 0.5y^4, \rho(y) = y^5$, the threshold increases by moving from $m = 1$ to $m = 2$, but after that it starts decreasing. For $\lambda(y) = y^2, \rho(y) = y^3$, the thresholds already start decreasing by moving from binary to an alphabet of size 4. We have observed for various other ensembles that if there are no degree 2 variable nodes then the threshold already starts decreasing by moving from binary to an alphabet of size 4.

## 4.2 Density Evolution for the Ensemble EGF $(\lambda, \rho, m)$

The ensemble EGF $(\lambda, \rho, m)$ is a subset of the ensemble EGL $(\lambda, \rho, m)$. We prove this and we characterize the set of messages for the ensemble EGF $(\lambda, \rho, m)$ in the following lemma.

$\lambda(y) = y, \rho(y) = y^2$
$\epsilon^{\text{sh}} \approx 0.6667$

| $m$ | $\epsilon^{\text{IT}}$ |
|---|---|
| 1 | 0.5 |
| 2 | 0.5775 |
| 3 | 0.6183 |
| 4 | 0.6369 |
| 5 | 0.6446 |
| 6 | 0.6464 |
| 7 | 0.6453 |
| 8 | 0.6425 |
| 15 | 0.616 |

$\lambda(y) = 0.5y + 0.5y^4, \rho(y) = y^5$
$\epsilon^{\text{sh}} \approx 0.4762$

| $m$ | $\epsilon^{\text{IT}}$ |
|---|---|
| 1 | 0.4 |
| 2 | 0.4487 |
| 3 | 0.4353 |
| 4 | 0.4194 |

$\lambda(y) = y^2,\ \rho(y) = y^3$
$\epsilon^{\text{sh}} = 0.75$

| $m$ | $\epsilon^{\text{IT}}$ |
|---|---|
| 1 | 0.6474 |
| 2 | 0.6348 |
| 3 | 0.6192 |

***Table 1:*** *Thresholds for the ensemble EGL $(\lambda, \rho, m)$ for various degree distributions.*

**Lemma 4.4 (EGF $(\lambda, \rho, m)$ as a subset of EGL $(\lambda, \rho, m)$)** *The mapping $f(\alpha) = \omega\alpha$, where $\omega \in \mathrm{GF}^*(2^m)$ and $\alpha \in \mathrm{GF}(2^m)$ is equivalent to a mapping $g(b) = Wb$, where $b \in \mathrm{GF}_2^m$ and $W \in \mathrm{GL}_2^m$. Hence all the messages in the belief propagation decoder are equivalent to the subspaces of the vector space $\mathrm{GF}_2^m$. In fact all the possible subspaces of the vector space $\mathrm{GF}_2^m$ do arise in the belief propagation decoder.*

For the derivation of density evolution equations, we observe that for $m \leq 3$, after the edge action a message of same dimension gets mapped to any other message of same dimension by equal number of mappings. Hence we can again combine the probability of the messages of the same dimension and do the density evolution over the dimension of the messages. Thus the density evolution equations for the ensemble EGF$(\lambda, \rho, m)$ is same as that of EGL$(\lambda, \rho, m)$ for $m \leq 3$ as given in Lemma 4.3. However for $m > 3$, it is no longer true that any message of the same dimension gets mapped to any other message of same dimension. The set of messages of given a dimension are partitioned into several orbits under the action of the label group. In order to accomplish density evolution we need to keep track of each orbit. This quickly becomes quite cumbersome. Also note that the performance of the ensemble *does* in general depend on the primitive element which one chooses to represent the field. I.e., two isomorphic fields do not in general yield identical equations. More precisely, if we run density evolution for a fixed number of iterations then the performance of two isomorphic fields is in general strictly different. An example is GF$(32)$ with fields defined with respect to the irreducible polynomials $1+z^3+z^5$ and $1+z+z^2+z^3+z^5$. The difference though between these two fields is very small and the resulting difference in the threshold is of the order of $10^{-4}$. Note also that within this precision, the ensembles EGF$(\lambda,\rho,m)$ and EGL$(\lambda,\rho,m)$ seem to have the (approximately) same threshold.

For the general case, an analysis in terms of density evolution is in principle possible but practically difficult. Even for codes over GF$(4)$ densities already "live" in $\mathbb{R}^3$. The BP threshold can be computed numerically by Monte Carlo methods in the same way as this is done in the setting of turbo codes, [17].

Slightly less ambitious, one can investigate the behavior of density evolution close to the desired fixed-point and derive a stability condition.

**Lemma 4.5 (Stability Condition for Non-Binary Ensembles)** *Consider the ensemble LDPC$(n,\lambda,\rho)$. Assume that transmission takes place over a BMS channel with L-density* $\mathsf{a}$ *and associated Battacharya constant* $\mathfrak{B}(\mathsf{a})$. *If*

$$\lambda'(0)\rho'(1)\frac{(1+\mathfrak{B}(\mathsf{a}))^m-1}{2^m-1} > 1, \tag{8}$$

*then the desired fixed-point is not stable.*

# 5 Upper Bound on MAP Thresholds

As pointed out in the introduction, the codes defined in this paper are linear. Therefore, if we assume transmission over a BMS channel and perform MAP decoding the (Generalized) EXIT curve fulfills the (Generalized) Area Theorem, [1, 11]. As was shown in [12, 14], the Area Theorem can be used to give an upper bound on the MAP threshold which is conjectured to be tight in a quite general setting.

We can proceed along exactly the same lines in our case. Let us start by defining the BP EXIT curve. Consider the ensemble EGL$(\lambda,\rho,m)$ where $(\lambda,\rho)$ and $m$ are fixed. Let us look at the asymptotic performance. More precisely, we first fix the number of iterations and let the block length tend to infinity. Now let the number of iterations tend to infinity. Let $\mathrm{P}^*(\epsilon,i)$ be the probability (in this limit) that the *extrinsic* decision of a random symbol contains $i$ degrees of freedom. I.e., $\mathrm{P}^*(\epsilon,0)$ is the probability that the symbol has been determined, $\mathrm{P}^*(\epsilon, 1)$ is the probability that the symbol is known to be one of possible two (and there is a uniform distribution), and so on. Then the EXIT curve under BP decoding is given by:

$$h^{\mathrm{BP}}(\epsilon) = \sum_{i=0}^{m} i\,\mathrm{P}^*(\epsilon, i).$$

Since the BP decoder is in general suboptimal we have

$$h^{\mathrm{MAP}}(\epsilon) \leq h^{\mathrm{BP}}(\epsilon),$$

where $h^{\mathrm{MAP}}(\epsilon)$ is the EXIT curve under MAP decoding. By the Area Theorem we have

$$\int_{\epsilon^{\mathrm{MAP}}}^{1} h^{\mathrm{MAP}}(\epsilon)\, d\epsilon = r_{\mathrm{as}}.$$

Hereby $\epsilon^{\mathrm{MAP}}$ is the MAP threshold of the ensemble and $r_{\mathrm{as}}$ is the average rate of the ensemble. In general $r_{\mathrm{as}} \geq r_{\mathrm{design}} = 1 - \frac{\int_0^1 \rho(x)dx}{\int_0^1 \lambda(x)dx}$ (for most "reasonable" ensembles we have equality, see [13]). We conclude that if for some $\overline{\epsilon}^{\mathrm{MAP}}$

$$\int_{\overline{\epsilon}^{\mathrm{MAP}}}^{1} h^{\mathrm{BP}}(\epsilon)\, d\epsilon = r_{\mathrm{design}},$$

then $\overline{\epsilon}^{\mathrm{MAP}} \geq \epsilon^{\mathrm{MAP}}$. In the binary case reference [13] gives some sufficient conditions for this bound to be tight. In short, the bound is tight if the residual graph which we get after running BP decoding at $\epsilon = \overline{\epsilon}^{\mathrm{MAP}}$ until the BP decoder is stuck contains only a single codeword with high probability. Fig. 3 shows several BP EXIT functions and gives the corresponding upper bounds on the MAP thresholds. These thresholds are conjectured to be tight. Note that these values of $\overline{\epsilon}_m^{\mathrm{MAP}}$ converge quickly to the Shannon threshold of $\frac{2}{3}$.

# 6 Conclusion

Following the lead of Gallager, Davey and MacKay, as well as Hu, we have investigate the performance of non-binary LDPC ensembles. In particular, assuming that transmission takes place over the BEC($\epsilon$), we have given a compact representation of the density evolution equations for the ensemble $\mathrm{EGL}(\lambda, \rho, m)$, we have derived the stability condition, and we have shown how to compute an upper bound on the MAP threshold via the area theorem.

In many ways this paper is only the beginning of a systematic investigation of non-binary iterative ensembles.

Let us state here what we consider to be some of the most interesting questions that remain unanswered. From the examples we have investigated, it seems that for a fix degree distribution the threshold is a unimodal function of the alphabet size. If there are sufficiently many degree-two variable nodes the threshold initially rises and eventually decays again as $m$ is increased. Otherwise it decrease right away. This is somewhat disappointing. Although, the underlying binary graph becomes denser and denser as $m$ increases (and in turn the MAP threshold converges to the Shannon limit very rapidly) the performance of the iterative decoder seems not to approach the Shannon limit.

There is one degree of freedom which was already suggested in [3] and which we have not considered so far. In all our analysis we assumed a uniform distribution on the edge labels. In out setting it is natural to allow a non-uniform distribution on the edge labels in such a way that the distribution respects the underlying algebraic structure. E.g., the ensemble $\text{EGF}(\lambda,\rho,m)$ can be considered a special case of the ensemble $\text{EGL}(\lambda,\rho,m)$ where we put a uniform distribution on the labels corresponding to field elements and zero weight on all other labels. Obviously there are many degrees of freedom that could be explored.

By a proper exploitation of these degrees of freedom one can hopefully find yet another way of approaching capacity, adding to our understanding of capacity approaching iterative coding schemes.

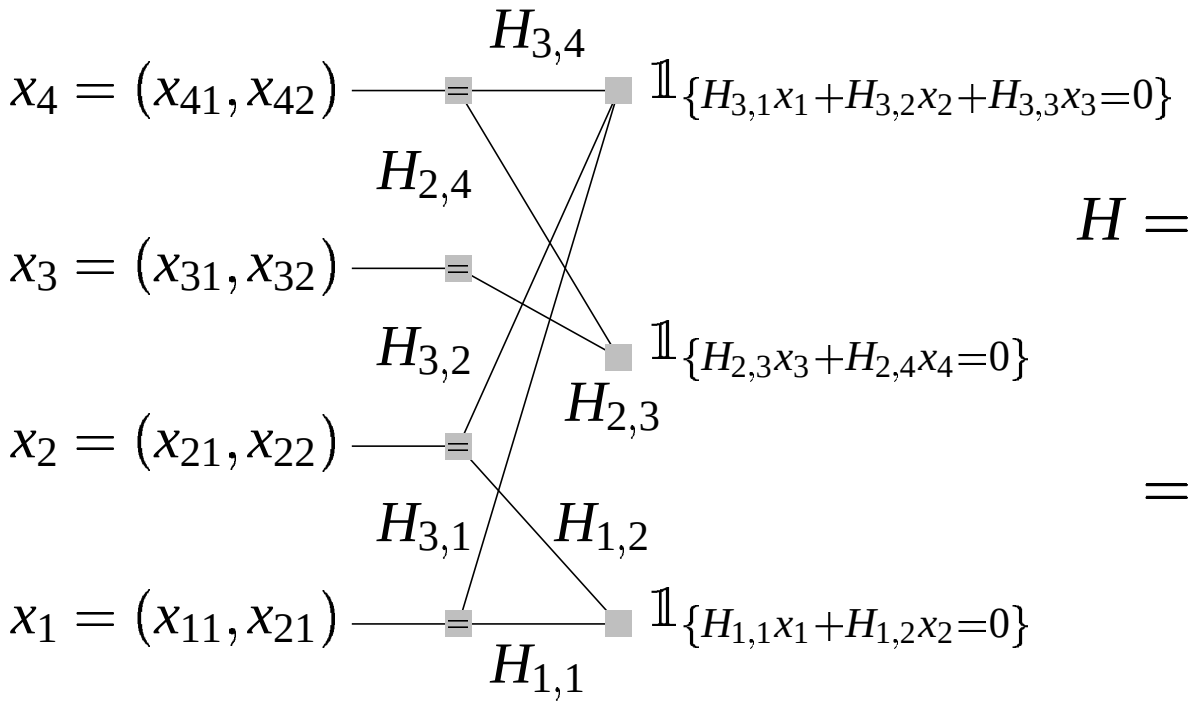

$$H = \begin{pmatrix} H_{1,1} & H_{1,2} & 0 & 0 \\ 0 & 0 & H_{2,3} & H_{2,4} \\ H_{3,1} & H_{3,2} & 0 & H_{3,4} \end{pmatrix} = \begin{pmatrix} 1 & 1+z & 0 & 0 \\ 0 & 0 & z & z \\ 1+z & z & 0 & 1 \end{pmatrix}$$

***Figure 1:*** *The FSFG of a simple code over* GF(4) *and its associated parity-check matrix* $H$. *The primitive polynomial generating* GF(4) *is* $p(z) = 1 + z + z^2$.

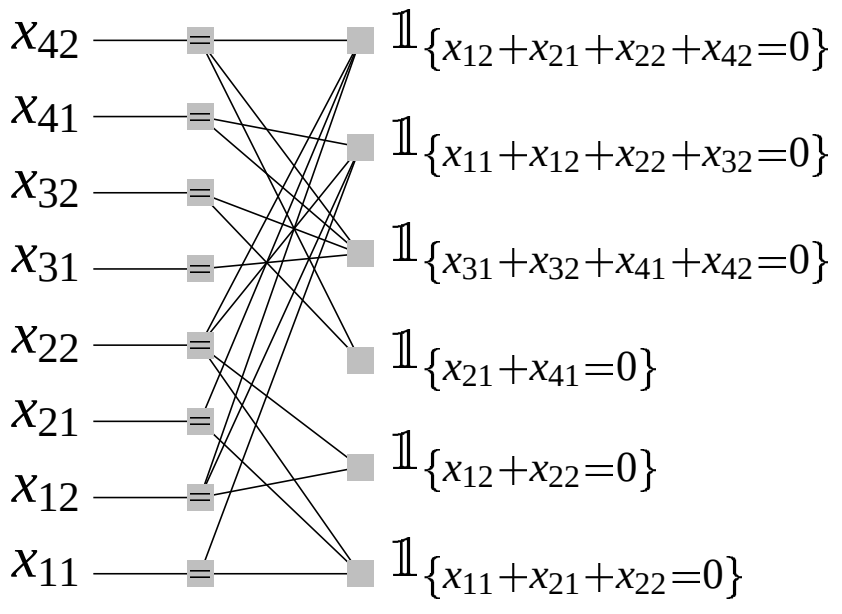

$$H_b = \begin{pmatrix} 1 & 0 & 1 & 1 & 0 & 0 & 0 & 0 \\ 0 & 1 & 0 & 1 & 0 & 0 & 0 & 0 \\ 0 & 0 & 0 & 0 & 0 & 1 & 0 & 1 \\ 0 & 0 & 0 & 0 & 1 & 1 & 1 & 1 \\ 1 & 1 & 0 & 1 & 0 & 0 & 1 & 0 \\ 0 & 1 & 1 & 1 & 0 & 0 & 0 & 1 \end{pmatrix}$$

***Figure 2:*** *The FSFG and its associated parity-check matrix corresponding to the equivalent binary code of the code given in Fig. 1.*

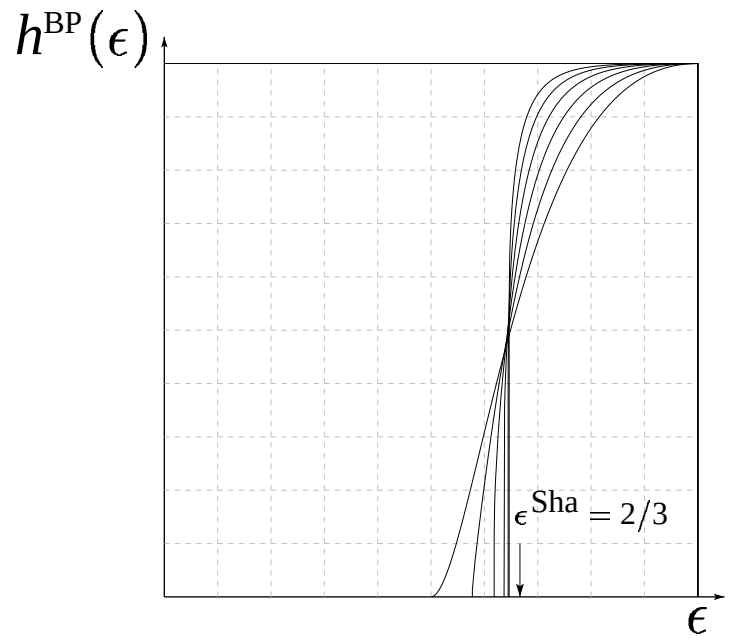

***Figure 3:*** *BP EXIT curves for the* $(2,3)$*-regular ensembles over* $\mathrm{GF}(2^m)$ *for* $m = 2, 3, 4, 5$ *and* 6 *and transmission over BEC*$(\epsilon)$. *By integrating the area under the respective BP EXIT curves starting from* $\epsilon = 1$ *until the area is equal to* $1/3$ *we get the following upper bounds on the MAP thresholds:* $\bar{\epsilon}^{MAP}_{m=1} = 0.5$, $\bar{\epsilon}^{MAP}_{m=2} \approx 0.5775$, $\bar{\epsilon}^{MAP}_{m=3} \approx 0.6209$, $\bar{\epsilon}^{MAP}_{m=4} \approx 0.6426$, $\bar{\epsilon}^{MAP}_{m=5} \approx 0.6540$, $\bar{\epsilon}^{MAP}_{m=6} \approx 0.6599$. *These values should be compared to the Shannon threshold of* $2/3 \approx 0.66667$. *These upper bounds are conjectured to be tight.*